\documentclass[10pt]{iopart}
\begin{document}
\title[Hamiltonian dynamics of extended...]
{Hamiltonian dynamics of extended objects: Regge-Teitelboim model}
\author{R  Capovilla\dag , A Escalante\dag,
J Guven\ddag  and E Rojas\P}
\address{\dag\
Departamento de F\'{\i}sica,
Centro de Investigaci\'on
y de Estudios Avanzados,
Apdo. Postal 14-740, 07000 M\'exico,
D. F.,
MEXICO}
\address{\ddag\
Instituto de Ciencias Nucleares,
Universidad Nacional Aut\'onoma de M\'exico,
Apdo. Postal 70-543, 04510 M\'exico, DF, MEXICO}
\address{\P\ Facultad de F\'{\i}sica e Inteligencia
Artificial,
Universidad Veracruzana, 91000 Xalapa, Veracruz, MEXICO}

\begin{abstract}
We consider the Regge-Teitelboim model for a relativistic extended
object embedded in a fixed background Minkowski spacetime, in which
the dynamics is determined by an action proportional to the integral
of the scalar curvature of the worldvolume spanned by the object in
its evolution. In appearance, this action resembles the
Einstein-Hilbert action for vacuum General Relativity: the equations
of motion for both are second order; the difference is that here the
dynamical variables are not the metric, but the embedding functions
of the worldvolume. We provide a novel Hamiltonian formulation for
this model. The Lagrangian, like that of General Relativity, is
linear in the acceleration of the extended object. As such, the
model is not a genuine higher derivative theory, a fact reflected in
the order of the equations of motion. Nevertheless, as we will show,
it is possible as well as useful to treat it as a `fake' higher
derivative system, enlarging the phase space appropriately. The
corresponding Hamiltonian on this phase space is constructed: it is
a polynomial. The complete set of constraints on the phase space is
identified. The fact that the equations of motion are of second
order in derivatives of the field variables manifests itself in the
Hamiltonian formulation through the appearance of additional
constraints, both primary and secondary. These new constraints are
second class. In this formulation, the Lagrange multipliers
implementing the primary constraints get identified as
accelerations. This is a generic feature of any Lagrangian linear in
the acceleration possessing reparametrization invariance.
\end{abstract}


\section{Introduction}

The idea that spacetime may be viewed as the trajectory of an
extended object, or brane, embedded in a background spacetime was
proposed by Regge and Teitelboim [RT] back in 1977 \cite{RT},
inspired by the Nambu-Goto [NG] model for relativistic strings
\cite{NG}.  Over the years, this idea has been revisited
\cite{DPR,Pavsic,GW,Tapia,Maia,DK,D,KD}. Regge and Teitelboim's
proposal was to  consider four-dimensional spacetime as the
worldvolume of some three-dimensional spacelike object embedded in a
fixed ten-dimensional flat Minkowski background; ten being the
minimum number of dimensions necessary to capture General Relativity [GR]
within this framework. The worldvolume is described by its embedding
functions into the background spacetime. The dynamics of the
extended object is determined by an action proportional to the
worldvolume integral of the worldvolume scalar curvature. If the
metric itself is the dynamical variable, this coincides with the
Einstein-Hilbert action for vacuum GR. In the
RT model, however, the dynamical variables are the embedding
functions themselves; the worldvolume metric is thus a derived
geometric quantity. This leads to important differences, both of a
technical and a conceptual nature, between the RT model and vacuum
GR. To consider a fixed background spacetime is anathema to many
relativists, but it is standard in string theory. For example, the
NG action is proportional to the area of the worldvolume spanned by
a relativistic string in a fixed background, the point of departure,
in its various forms, of string theory \cite{Polchinski}.

Besides the fact that the RT action is what would occur naturally to
a relativist when looking at relativistic extended objects---indeed
the title of the RT paper is {\it Gravity \'a la String}---the RT
action possesses a key feature in common with the NG action: they
both lead to equations of motion that are of  {\it second order} in
derivatives of the field variables, the embedding functions.  While
this is obvious for the NG action, it is not for the RT action.
Among candidate actions for extended objects that are local,
reparametrization invariant and invariant under rigid motions of the
ambient spacetime they are quite special in this
sense\footnote{Another model that shares this property is the Tolman
model, the worldvolume integral of the mean extrinsic curvature,
which exists however only for hypersurfaces.} --most actions one can
construct which satisfy these requirements in general will involve
derivatives higher than second; this is the case, for instance, for
the action proportional to the integral of the squared mean
curvature proposed by Polyakov as an effective model in QCD
\cite{Polyakov}.

An additional motivation for re-examining the RT action comes from
recent developments in brane world scenarios \cite{Nima,RS}. It
should be emphasized, however, that such scenarios are  much more
complicated, as they involve a non-trivial background spacetime.
With this motivation in mind, Karasik and Davidson recently proposed
a Hamiltonian formulation of the RT model \cite{KD}.  With the
addition of a Lagrange multiplier, they  adapt the ADM canonical
formulation of GR to the RT model, so that it is written as a system
that depends only on the velocities of the extended object. The
price they pay is that the Legendre transformation yielding the
Hamiltonian is {\it not} simple. Moreover, the explicit form of the
Hamiltonian, as well as the constraints that generate
reparametrization invariance, are  not polynomials in the phase
space variables. Despite these difficulties, they proceed to include
possible matter fields, they consider both canonical and path
integral quantizations of the system, and analyze a minisuperspace
model.

In this paper we provide an alternative  Hamiltonian formulation of
the RT model yielding a Hamiltonian functional that is {\it
polynomial} in the phase space variables. How this is accomplished
is by extending the phase space to include, in addition to the
position of the brane at a fixed time and its conjugate momentum,
also the velocity of the brane at a fixed time and its conjugate
momentum. In other words, we treat the RT model as a higher
derivative theory. This may seem perverse, in light of what we said
above: the RT model is {\it not} a higher derivative theory. Its
action does, however, depend on the brane acceleration; why the
model is not a higher derivative one is because the dependence on
the acceleration is {\it linear}. As we will show, this means that
the extra degrees of freedom we introduce are pure gauge, and thus
harmless. We will rely heavily on the formalism developed in
\cite{hambranes} for truly higher derivative models of relativistic
extended objects. The simplification of the functional form of the
Hamiltonian comes with a price. As was to be expected from
reparametrization invariance, there are constraints. We find that
there are second class constraints and that the constraint algebra
is quite complicated. Nevertheless, we do possess at our disposal
Dirac's powerful algorithm to deal with this difficulty in a
systematic way (see {\it e.g.} \cite{HT}). Even if the model itself
is not completely compelling, it does provide an excellent test-bed
for the Dirac framework within the context of higher derivative
theories.

The paper is organized in the following way: In Sect. 2, we
introduce some basic notions of the worldvolume geometry, and the RT
action together with its equations of motion. We also recall how the
equations of motion can be written as a conservation laws in terms
of a stress-tensor, and its relation with the conserved linear
momentum associated with invariance under background translation. In
Sect. 3, we perform a $d+1$ decomposition of the worldvolume
geometry, in the spirit of the ADM canonical formulation of GR. This
involves focusing on the geometry of the brane $\Sigma$. The $d+1$
decomposition of the RT action allows us to identify the appropriate
Lagrangian functional for the model. In Sect. 4, we construct the
Hamiltonian formulation of the model. Firstly, we  obtain the
appropriate phase space. Secondly, via a Legendre transformation, we
arrive at the canonical Hamiltonian. Next, we identify the primary
and secondary constraints associated with the reparametrization
invariance of the model and we discuss their content. Finally, we
look briefly at Hamilton's equations.
We conclude in Sect. 5 with a few brief remarks.

\section{Regge-Teitelboim model}

We consider the $(d+1)$-dimensional worldvolume $m$ spanned by the
evolution of a $d-$dimensional spacelike extended object, or brane,
$\Sigma$ in a fixed Minkowski background spacetime of dimension $D$.
The worldvolume $m$ is described by the embedding
$x^{\mu}=X^{\mu}(\xi^{a})$, where $x^{\mu}$ are local coordinates
for the background spacetime, $\xi^{a}$  local coordinates for the
worldvolume $m$, and $X^\mu$ the embedding functions ($\mu ,\nu=
0,1, \ldots, N-1$; $a,b=0,1,\ldots,d$). We denote by $X^\mu_a =
\partial_a X^\mu = \partial X^\mu / \partial \xi^a$ the tangent
vectors to $m$. Their inner product gives the induced metric on $m$,
$g_{ab} = \eta_{\mu\nu} \, X^\mu_a X^\nu_b = X_a \cdot X_b$, where
$\eta_{\mu\nu}$ is the Minkowski metric with one minus sign. (To
avoid notational clutter, wherever possible we will not write the
spacetime indices.) We denote by $g^{ab}$ the inverse of $g_{ab}$.
The $D-d-1$ unit spacelike normals $n^\mu_i$ to $m$ are defined
implicitly by $n_i \cdot X_a = 0$, $n_i \cdot n_j = \delta_{ij}$ ($i
, j = 1, \ldots, D-d-1$). The extrinsic curvature of $m$ is
$K_{ab}{}^i = - n^i \cdot \partial_a X_b$, and the mean curvature is
its trace $K^i = g^{ab} K_{ab}{}^i$. The scalar curvature ${\cal R}$
of $m$ can be obtained either directly from the induced metric
$g_{ab}$, or, in terms of the extrinsic curvature, via the
contracted Gauss-Codazzi equation:
\begin{equation}
{\cal R}= K^{i}K_i - K_{ab}^{i}K^{ab}_i\,.
\label{eq:GC}
\end{equation}

The RT action for a $d-$dimensional brane is given by
\begin{equation}
S _{RT} [X^\mu ]= \frac{\alpha}{2} \int_m d^{d+1}\xi \,\sqrt{-g}\,{\cal R}\,,
\label{eq:action}
\end{equation}
where the constant $\alpha$ has dimensions $ [L]^{(1-d)}$. Note that
for a relativistic string (with $d=1$), this action is a topological
invariant because of the Gauss-Bonnet theorem, so that it has empty
equations of motion.

The equations of motion that follow from this action are \cite{RT}
\begin{equation}
\alpha \; {\cal G}^{ab} \; K_{ab}{}^i = 0\,,
\label{eq:eom}
\end{equation}
where ${\cal G}_{ab} = {\cal R}_{ab} - (1/2) {\cal R} g_{ab}$ is the
worldvolume Einstein tensor, with ${\cal R}_{ab}$ the Ricci tensor.
These equations of motion are of second order in derivatives of the
embedding functions because of the presence of the extrinsic
curvature. There are only $D-d-1$ equations, along the normals; the
remaining $d+1$ tangential components are satisfied identically, as
a consequence of the reparametrization invariance of the action. If
$d=3$ and $D=10$, there are six equations: this is the same as the
number of Einstein equations for a four-dimensional spacetime modulo
the Bianchi identities.

It is of interest  to compare the RT model to the NG model,
proportional to the area of the worldvolume spanned by the extended object in its
evolution:
\begin{equation}
S_{NG}  [X^\mu ]= - \mu \int_m d^{d+1}\xi \,\sqrt{-g}\,,
\label{eq:actionng}
\end{equation}
with equations of motion
\begin{equation}
\mu \; g^{ab} \; K_{ab}{}^i = 0\,,
\label{eq:eomng}
\end{equation}
the mean curvature vanishes. Comparison with the RT equations of motion shows
that  the Einstein tensor plays the same role there as here the induced metric.

The equations of motion (\ref{eq:eom}) can be written in terms of a stress-tensor $f^{\mu a}$ as \cite{RT,Noether}
\begin{equation}
\alpha \; {\cal G}^{ab} \; K_{ab}{}^i n^\mu_i  =
\nabla_a f^{\mu a}\,,
\label{eq:conse}
\end{equation}
where the stress tensor is given by
\begin{equation}
f^{\mu a} = - \alpha \; {\cal G}^{ab} X^\mu_b\,,
\label{eq:stress}
\end{equation}
and (\ref{eq:conse}) follows from the Bianchi identity $\nabla_a {\cal G}^{ab} = 0$ and the
definition of the extrinsic curvature, $\nabla_a X_b = - K_{ab}{}^i  \; n_i$. The stress-tensor
is purely tangential, and this is related to the fact that the equations
of motion are of second order in derivatives of the embedding functions.
Moreover, if we let $\eta^a$ denote the timelike unit normal from $\Sigma$ into $m$,
then we can construct the quantity
\begin{equation}
\pi^\mu =  \eta_a f^{\mu a}= - \alpha \;\eta_a \; {\cal G}^{ab} X^\mu_b\,,
\label{eq:mom}
\end{equation}
which is the conserved linear momentum density associated with
the invariance of the action (\ref{eq:action}) with respect to background translations
\cite{Noether}.

\section{ADM decomposition}

We split the $d+1$ worldvolume coordinates $\xi^a$ into an arbitrary
evolution parameter $t$
and $d$ coordinates $u^A$ that parametrize, at fixed $t$, the spacelike brane $\Sigma$
($A,B,\dots= 1, \dots, d$), in the sense that $\Sigma$ is represented by the
embedding $x^\mu = X^\mu (t=$const.$, u^A$). Alternatively, $\Sigma$ can
be described by its
embedding in the worldvolume itself, $\xi^a = X^a (u^A)$. These two descriptions are
related by composition. (Details can be found in \cite{hambranes}).)
The tangent vectors to $\Sigma$ are denoted by $X^\mu_A = \partial X^\mu
/ \partial u^A $, and the induced metric on $\Sigma$ is $h_{AB} =
X_A \cdot X_B$, with inverse $h^{AB}$, and determinant $h$. We denote with
${\cal D}_A$ the $\Sigma$ covariant derivative
compatible with $h_{AB}$. The unit timelike normal to $\Sigma$ into $m$ is $\eta^\mu$, defined
by $\eta \cdot X_A = 0$, $ \eta \cdot \eta = - 1$.

The velocity $\dot{X} = \partial_t X $ is a vector tangent to the worldvolume $m$.
It can be expanded in components with respect to the worldvolume basis $\{ \eta , X_A \}$
as
\begin{equation}
\dot{X} = N \, \eta + N^A \, X_A\,,
\end{equation}
where $N$ and $N^A$ are the lapse function and the shift vector, respectively.

We decompose now the geometry of the worldvolume along the basis $\{ \dot{X} , X_A \}$,
adapted to the evolution of $\Sigma$.
The decomposition of the induced metric $g_{ab}$ with respect to this
basis is
\begin{equation}
g_{ab}= \left(
\begin{array}{cc}
(-N^2 + N^AN^Bh_{AB}) &N^Bh_{AB} \\
N^A h_{AB} & h_{AB}
\end{array}
\right) \,,
\end{equation}
and it follows that its determinant takes the simple form
\begin{equation}
g = -N^2 h\,.
\end{equation}
We will also need the decomposition
of the inverse metric:
\begin{equation}
g^{ab}= \frac{1}{N^2}\left(
\begin{array}{cc}
-1 &N^A \\
N^B & (N^2h^{AB} - N^AN^B)
\label{eq:gidec}
\end{array}
\right) \,.
\end{equation}
For the extrinsic curvature tensor, we have
\begin{equation}
K_{ab}{}^i = - \left(
\begin{array}{cc}
 n^i \cdot \ddot{X}
 & n^i \cdot {\cal D}_A \dot{X}
 \\
n^i \cdot {\cal D}_B \dot{X}  & n^i \cdot {\cal D}_A {\cal D}_B X
\end{array}
\right) \,.
\label{eq:exdec}
\end{equation}

In order to find the decomposition of the scalar curvature ${\cal R}$, we
use the decompositions (\ref{eq:gidec}), (\ref{eq:exdec}), and the Gauss-Codazzi
equation (\ref{eq:GC}). We find
\begin{equation}
{\cal R} = \frac{1}{N^2} \left[ - 2
(n^i \cdot {\cal D}^A {\cal D}_A X )
  (n_i \cdot \ddot{X}) + 2 J_{\cal R} \right]\,,
\end{equation}
where we have isolated the part of ${\cal R}$ that does not depend on the acceleration  $\ddot{X}$:
\begin{eqnarray}
\fl
{\cal J}_{\cal R} ( X, \dot{X} ) &=& {1 \over N^2} [
     h^{AB} ( n^i \cdot {\cal D}_A \dot{X}) ( n_i \cdot {\cal D}_B \dot{X})
+ 4 N^{[C} h^{B]A}  (n^i \cdot  {\cal D}_A {\cal D}_B X )
(  n_i \cdot {\cal D}_C \dot{X} )
\nonumber \\
\fl
 &+&     2 ( h^{A[C} N^{B]} N^D + h^{A[B} h^{C]D} ) ( n^i \cdot {\cal
D}_A {\cal D}_B X)
( n_i \cdot {\cal D}_C {\cal D}_D X) ]\,.
\label{eq:jr}
\end{eqnarray}

We are now in a  position to rewrite  the RT action
(\ref{eq:action}) in terms of quantities defined with respect to the
brane $\Sigma$ as
\begin{equation}
S_{RT} [X^\mu] = \int  dt \, L [X,\dot{X},\ddot{X}] \,,
\label{eq:action1}
\end{equation}
where we identify the Lagrangian functional
\begin{eqnarray}
L [X,\dot{X},\ddot{X}] &=& \int_{\Sigma}d^d u \, 
{\cal L} (X,\dot{X},\ddot{X}) \,,
\nonumber
\\
&=&\int_\Sigma d^d u \; \frac{\alpha \sqrt{h}}{N}\,
\left[ -
(n^i \cdot {\cal D}^A {\cal D}_A X )
  (n_i \cdot \ddot{X})  + \, {\cal J}_{\cal R} \right]\,,
\label{eq:lagrangian}
\end{eqnarray}
and  ${\cal L}(X,\dot{X},\ddot{X})$ denotes the Lagrangian density.
The important thing to note is that the dependence of the Lagrangian
on the acceleration $\ddot{X}$ is linear. The dependence on the
velocity $\dot{X}$ is somewhat complicated: it enters through the
lapse function
   $N$, the normal vectors
$n^i$, as well as the quantity ${\cal J}_{\cal R}$. We note that the
dependence on the position vectors $X$ comes in only through its
derivatives with respect to the coordinates on $\Sigma$, $u^A$.

\section{Hamiltonian formulation}

The Lagrangian functional (\ref{eq:lagrangian}) is the starting
point of the Hamiltonian formulation of the model defined by the
action (\ref{eq:action}). Although we know from the outset that the
model is not a higher derivative theory, we will treat it as though
it were. As will become clear below, although additional degrees of
freedom are introduced, there is no inconsistency in the strategy:
the acceleration does not need to be specified as an initial
condition.

The phase space involves two conjugate pairs $\{ X,p; \dot{X}, P
\}$, where $p$ and $P$ denote the momenta conjugate to the position
functions  $X$ and  the velocities $\dot{X}$, respectively. They are
defined by
\begin{eqnarray}
P &=&
{\delta L \over \delta \ddot{X}}\,, \\
p &=& {\delta L \over \delta \dot{X}} - \partial_t \left(
{\delta L \over \delta \ddot{X}}\right)\,.
\end{eqnarray}
For the momenta $P$ we obtain immediately from
(\ref{eq:lagrangian}):
\begin{equation}
P = - \alpha \; {\sqrt{h} \over N} \, \left( n_i \cdot
{\cal D}_A{\cal D}^A X \right)\,n{}^i\,.
\label{eq:P}
\end{equation}
They are normal to the worldvolume. Note that the right hand side is
a function only of $X$ and $\dot{X}$; it does not involve $\ddot
{X}$. The factor of $\sqrt{h}$ tells us that the momenta  are
spatial densities. For $p$ we obtain \cite{hambranes}
\begin{equation}
p = \sqrt{h} \; \pi
+ \partial_A \left[ N^A P - \alpha \sqrt{h}  \left( n_i \cdot
{\cal D}^A \dot{X} \right)
\,n^i  \right]\,,
\label{eq:p}
\end{equation}
where $\pi$ is the conserved linear momentum density  obtained from
Noether's theorem (\ref{eq:mom}). $p$ and $\pi$ differ only by a
spatial divergence; their integral over a closed geometry is the
same.

We have thus identified the appropriate phase space for the RT model. We emphasize
that a truly higher derivative model would have qualitatively different momenta: $P$ would
continue to be normal to the worldvolume but it would depend on the acceleration $\ddot{X}$;
the $\pi$ contribution to $p$ would possess a part normal to the worldvolume and $p$ would
depend on $X$ with three dots.

The canonical Hamiltonian is obtained via a  Legendre transformation of the Lagrangian functional
(\ref{eq:lagrangian}) with respect to both $X$ and $\dot{X}$ (see {\it e.g.} \cite{hambranes}):
\begin{equation}
H_0 [X,p;\dot{X},P]=\int_\Sigma d^d u \,\left( p\cdot \dot{X}
+ P \cdot \ddot{X} \right) - L [X,\dot{X},\ddot{X}]\,.
\end{equation}
Since the Lagrangian is linear in the acceleration, and the term
$p \cdot \dot{X}$ is already in canonical form, we obtain immediately
\begin{equation}
H_0 [X,p;\dot{X}] = \int_\Sigma d^A u \,\left[ p \cdot \dot{X}
 - \alpha\frac{\sqrt{h}}{N}\, {\cal J}_{\cal R} (X , \dot{X} ) \right]\,,
\label{eq:h0}
\end{equation}
where the quantity   ${\cal J}_{\cal R} (X , \dot{X} )$ is given by (\ref{eq:jr}).
Note that this Hamiltonian is independent of the higher momentum $P$,
linear in $p$, that comes in only in the first term, $ p\cdot \dot{X}$, and highly non-linear
in its dependence on the canonical variables $\dot{X}, X $.

Since the Lagrangian (\ref{eq:lagrangian}) is linear in the acceleration, its Hessian
vanishes identically,
\begin{equation}
{\cal H}_{\mu\nu} =  {  \delta L [X,\dot{X},\ddot{X}] \over \delta \ddot{X}^\mu  \delta \ddot{X}^\nu}  = 0\,.
\end{equation}
This implies, as expected,  that there are constraints on the phase space variables. The first set of (primary) constraints
is readily identified as the definition  of the higher momentum $P$, as given by (\ref{eq:P}). Therefore we have the $D$ primary constraints
\begin{equation}
C = P + \frac{\sqrt{h}}{N}\left( n_i \cdot
{\cal D}_A{\cal D}^A X\right) \,n^i = 0 \,.
\label{eq:pT}
\end{equation}
A word of caution is perhaps useful:
it is tempting to square this constraint, just as one does in the Hamiltonian formulation
of the NG model (see {\it e.g.} \cite{ADM}).  However, as argued convincingly by  Nesterenko in his Hamiltonian
formulation of a geometrical model of a relativistic particle, this
would lead to error   \cite{Nesterenko}. One would be throwing away $D-1$ primary constraints.

The total Hamiltonian that generates evolution in the phase space
is given by adding to
the canonical Hamiltonian (\ref{eq:h0}) the primary constraints (\ref{eq:pT}). This results in
the Hamiltonian:
\begin{equation}
H[X,p;\dot{X},P] = H_0 [X,p;\dot{X},P] +
\int_\Sigma d^d u \; \lambda \cdot C \,.
\label{eq:HTT}
\end{equation}
where $\lambda $ are Lagrange multipliers  enforcing
the constraints.

At this point, let us recall that the Poisson bracket appropriate for a
higher derivative theory is,
for two arbitrary phase space functionals, $f$ and $g$,
\begin{equation}
\left\lbrace f,g\right\rbrace = \int_\Sigma \left[
\frac{\delta f}{\delta X}\cdot
\frac{\delta g}{\delta p} +
\frac{\delta f}{\delta \dot{X}}\cdot
\frac{\delta g}{\delta P}
-  (f \leftrightarrow g) \right]\,,
\end{equation}
and that the time derivative of any phase space function is given by its
Poisson bracket with the total Hamiltonian
(\ref{eq:HTT}):
\begin{equation}
\partial_t f = \dot{f} = \left\lbrace f, H \right\rbrace \,.
\end{equation}

We have identified the $D$ primary constraints
(\ref{eq:pT}). This is not the whole story, however.  Consistency requires that
their conservation in time vanishes as well. In this case, this produces a set of secondary constraints. Setting
$H_0 = \int_\Sigma d^d u \; {\cal H}_0$, we find the $D$ secondary constraints
\begin{eqnarray}
{\cal S}_0 &=&
{\cal H}_0 = 0\,, \label{eq:s1}
\\
{\cal S}_A &=& p \cdot X_A + P \cdot \partial_A \dot{X} = 0\,, \label{eq:s2}
\\
{\cal S}_i &=& p \cdot n_i - n_i \cdot
\partial_A \left[ N^A P  - \alpha \sqrt{h}  \left( n_j \cdot
{\cal D}^A \dot{X} \right)
\,n^j  \right]\,.
\label{eq:s3}
\end{eqnarray}
There are no other, tertiary constraints. The first secondary
constraint (\ref{eq:s1}) is the vanishing of the canonical
Hamiltonian, the hallmark of reparametrization invariance. It
generates diffeomorphisms out of the extended object $\Sigma$ onto
the worldvolume $m$, and not necessarily normal to it. The second
constraint (\ref{eq:s2}) is the generator of diffeomorphisms
tangential to $\Sigma$. The structure of these two constraints is
the same found in the Hamiltonian analysis of genuine higher
derivative brane models in \cite{hambranes}. The third constraint
(\ref{eq:s3}) is, however, characteristic of a model linear in
acceleration; it expresses the fact that the momentum density
(\ref{eq:mom}) is tangential just as it is in the case of the NG
model. In general, if the action depends on accelerations in a
non-linear way, the momentum picks up a normal component
\cite{hambranes}.

Let us consider briefly the constraint algebra. (A detailed treatment is outside the
scope of this paper and will be considered
elsewhere \cite{ER}.) For this, it
is convenient to project the primary constraints (\ref{eq:pT}) along the basis
$\{ \dot{X} , X_A , n^i \} $ to obtain the equivalent set of constraints:
\begin{eqnarray}
C_0  &=& P \cdot \dot{X} = 0\,, \label{eq:p1} \\
C_A &=& P \cdot X_A  = 0\,, \label{eq:p2} \\
C_i  &=&  P \cdot n_i  +  \frac{\sqrt{h}}{N}\left( n_i \cdot
{\cal D}_A{\cal D}^A X\right)  = 0 \,. \label{eq:p3}
\end{eqnarray}
Again, the first two constraints are the same found in
\cite{hambranes} for truly higher derivatives brane models, the
third is particular to this model. A lenghty computation shows that
the constraints $\{ {\cal S}_0 , {\cal S}_A , {\cal C}_0, {\cal C}_A
\}$ are first class, {\it i.e.} in involution among themselves; the
constraints $\{ {\cal S}_i , {\cal C}_i \}$, however, are second
class.

The counting of physical degrees of freedom goes as follows: number = $(1/2)$
[ (dimension of the phase space) - 2 $\times$ (number of first class constraints) -
number of second class constraints) ]. In this case we obtain: $(1/2)
[ 4 D - 4 (d + 1) - 2 (D-d-1)] = D-d-1$: one physical degree of freedom along each normal,
just as is the case for the NG model \cite{ADM}.

Let us consider briefly the structure of the Hamilton's equations
that follow from the Hamiltonian (\ref{eq:HTT}). We will not write
them down explicitly since their form is complicated and not
particularly illuminating. In any case, one can find the needed
variational tools in \cite{hambranes}. The  first equation is
apparently a trivial identity
\begin{equation}
\partial_t{X} = \frac{\delta H}{\delta p} = \dot{X}\,,
\end{equation}
since the only dependence on $p$ in the Hamiltonian is through the
linear term $p\cdot \dot{X}$ in (\ref{eq:HTT}); however, this
equation serves to identify the canonical variable $\dot{X}$ with
the time derivative of the position functions $X$. The second
Hamilton equation is given by
\begin{equation}
\partial_t{\dot{X}} = \ddot{X} =  \frac{\delta H}{\delta P} = \lambda\,,
\end{equation}
and it identifies the Lagrange multipliers $\lambda$ with the acceleration $\ddot{X}$.
This is the hallmark of a theory linear in acceleration. One can verify that is
completely analogous to the situation one encounters when the dynamics
of a non-relativistic free particle is formulated using  the
Lagrangian $L(x,\ddot x) = - x\ddot x/2$, which is equivalent
to the quadratic in velocities form, modulo a boundary term.
In other words, despite appearances $\ddot X$ does not need to be specified among the
initial data which, happily, is consistent with the fact that the equations of motion are of second order in derivatives of the field variables.  The third Hamilton equation is
\begin{equation}
\partial_t  P = \dot{P}  = - {\delta H \over \delta \dot{X}}\,,
\end{equation}
and its role is to identify the form of the momenta
$p$ in terms of $X, \dot{X}, P$ and $\dot{P}$. Using the constraints, it reproduces (\ref{eq:p}).
Finally, the fourth Hamilton equation is
\begin{equation}
\partial_t p = \dot{p}= - {\delta H \over \delta X}\,,
\end{equation}
and it is the equation of motion (\ref{eq:eom}) in its canonical disguise.

\section{Concluding remarks}

In this paper we have presented a Hamiltonian formulation  of the RT
model for an extended object,  treated as a `fake' higher derivative
theory. Within our framework, the Hamiltonian function that
determines evolution is polynomial in the phase space variables. We
have identified the phase space constraints: there is a set of first
class constraints, associated with the symmetry of reparametrization
invariance, and a set of secondary constraints. One is now in a
position to  canonically quantize the RT model based on this
Hamiltonian formulation. It would be interesting to compare the
result with the formulation of the theory described by Karasik and
Davidson in \cite{KD}. In our formulation, the canonical
quantization of the model will require the secondary constraints to
be analyzed using the Dirac algorithm \cite{HT}. Work on this is in
progress. Finally, we note also that our formulation is not changed
substantially by the addition of a cosmological constant term and/or
matter fields. Terms get added to the momenta and the canonical
Hamiltonian in a way which leaves the structure unchanged.

\ack

We acknowledge partial support from CONACyT under
grant 44974-F as well as from DGAPA under grant IN119206. ER
acknowledges partial support from CONACyT
under grant C01-41639 and PRO\-MEP 2004-2007.

\end{document}